# Discussion Chain

Tomohiro Nobeyama

[Abstract]

**To make scientific truth more reliable and qualified, I propose to focus on the chain of discussion in a scientific journal rather than on each original paper. The value, quality and reliability should be judged by the form of the whole discussion chain, not only by an original report. Funding and sponsorship should also give more priority to the extension of the discussion chain. Maintaining the discussion chain and evaluating each scientific truth by the whole chain is the next value of the scientific community in the post-generative AI era.**

[Main text]

What is *scientific truth*? A paper published in the highest impact factor journal? An argument by the highest authority? No scientist would agree. Science is the work of building a stack of discussion, supported by reliable data or logic to the fullest extent possible, and a series of attempts to edge closer to the truth we recognize. Alternatively, science is driven by a group of people interested in passing down a "chain of discussion" from one generation to the next. It is time to return to this foundational agreement. To get better future, we have to conduct better research. And, to do so, to reckon what scientific truth is would be necessary for build a good future. I proposed that we should put more priority to make a "discussion chain", a whole discussion log on each scientific provocation and more value and rewards to construct and evaluate the shape of discussion chain as same as report of root scientific opinion itself. The value of a root opinion should be determined by the shape of discussion chain, not on the mere one-dimensional metrics such as citation number, journal impact factor. The shape includes how much critics scientific community can answer in what range around a root scientific opinion.

The length of the chain of discussion determines the strength of scientific truth. For example, Newtonian physics is widely accepted as a truth model. The authenticity of Issac Newton or the authority of Cambridge does not bring about confidence in this truth. After the publication of the *Principia*, scientists from various fields critiqued its contents and classical physics was shaped through the process of criticism. A solid consensus has been reached regarding its various applications and limitations. The establishment of these limitations is a source of acceptance.

Thus, we can now use the physics supported by the discussion chain in a way that solves problems in correct ranges of scope. If a problem is beyond the scope of the physics, we can use a different physical theory such as quantum physics.

Another example is the electronic theory of organic reactions, which explains organic reactions using the polarity and flow of electrons (represented by curled arrows)[1]. Of course, this method could not well explain the overlapping of molecular orbitals, which is the true mechanism of bond formation/breakage; more valuable methods are also used such as frontier orbital theory, *ab initio* quantum calculation, and other calculation-driven approaches[2]. However, the accumulated discussion about organic reactions has produced a cohesive tacit knowledge of the scope of validity of electron-pushing formalism so that organic chemists and biologists can reliably communicate with low-cost interpretation. The one still constitutes a major part of the chemistry curriculum[3]. In both examples, many critiques were organized as a sequential chain among scientists, revealing how much we rely on in what range in the world.

Writing a research paper is an act to elongate the discussion chain. For example, the introduction and discussion sections are drafted to link to the existing discussion chain in one's responsibility, and the results and the methods sections are drafted to create new links in the discussion chain. Each link adds to a discussion chain in a given field, and the readers of the manuscript can then evaluate the discussion chain. This is a critical aspect of publishing a paper. Peer review is regarded as a one-time elongation of discussion chain before showcasing your paper, rather than screening or gatekeeping.

However, people should not consider something as a scientific truth without a long and sufficient discussion chain, even in the case where the original report is indeed accurate. The critical trials of a new drug would be a good example. We can rely on the results of the trial because of the accumulated discussion chains of fundamental research papers from *in vitro*, *in silico*, and *in vivo* preclinical trials, sound statistical designs, and phases I and II clinical trials. In cases where a large-scale human experiment is conducted without such precedent, its results are less trustworthy

The importance of discussion chain–based evaluation emerges when one considers the fact that researchers can "fail" their experiments without evidence in the lab notes whether the log is accurate or not. One example is a paper about the cyclopropanation reaction of an inactive alkene by Franzen and Wittig[4,5]. Despite careful attempts to reproduce the results with Wittig, who is a historical organic chemist and one of the authors of Franzen et al.[4], they failed to reproduce the same results; thus, their first paper was retracted. Approximately five decades later, Chan et al.

observed the same cyclopropanation and they were surprised that the original finding was true[6] (The reproduction team asked for an independent reproduction of their results by W. Leitner.). The indetectable nickel ions—at least for Franzen at that time—were a key factor in the reaction. I am unaware whether Wittig used a different batch of reagents and/or if nickel ions went into the flask derived from the procedures by Franzen. It is just a little matter.

For spectators, only a discussion chain of a sufficient length would allow them to judge and recognize whether experiments are scientifically true or not. At the time the first paper by Franzen et al. was retracted, the discussion chain was short and only two connected links were present: the original report and comments on the reproducibility of their results. The extended chain of discussion has allowed us to move further toward the truth. Just like falsifiability, this is an important aspect of science.

This chain-based determination of truth is apparently similar to a block chain system[7]. Both systems rely on a chain of information, not simply on individual authenticity, such as banking, government identification, and personal credit. These systems are similar in that they require the entire chain to be available for an audience to verify, and the longer the chain is extended by the participants, the better truth is maintained However, a block chain recognizes any tree node that has an invalid hash; however, the discussion chain in science lets us judge the truth range and tendency on the structure and quality of the whole chain.

In an era of post-truth and generative AI, scientists will be working to elongate and evaluate the discussion chain.   For example, if you want to check whether a proposition is true, you will try to create a discussion chain and, until the chain becomes long and solid, you will not regard the fact as scientific truth, although it is published in a top journal or supported by some authority. To rephrase, we shift toward a study of the discussion chain, rather than a study of facts and focus on the evaluation of the chain itself.

One good way to create a discussion chain is to use a critique-based method. We challenge the object of discussion with an assumption and test whether the argument holds from various viewpoints. Of course, if a disadvantageous assumption surpasses a criterion, an argument will not be held. Considering the criterion, we proceed to check the validity of the assumption. If the assumption is ridiculous, we can judge that the argument is scientifically true. If not, we must look at the shape of one along with the tendency. Notably, the elongation of a discussion chain must have equality. Therefore, for instance, after you ask whether A is the truth, you also ask whether the proposition A is NOT the truth. The swing-style discussion must be maintained. The above example of the catalytic reaction demonstrates the importance of the swing-style discussion to elongate the discussion chain.

From the perspective of finding an algorithm to uncover scientific truth, for example, academic writing style would be quick to focus on the extensibility of the discussion chain among scientist rather than the preparation of a brilliant and impeccable story. Comment papers that are linked to original articles, such as letters to the editor, would probably be given higher value and admiration by the authors. Their original paper would not be seen as sufficiently true without a sufficient discussion chain; therefore, all authors would be apt to welcome critiques paper, reproduction paper, comment, letter to editor from others as current researchers prefer publishing their paper in peer-reviewed journals. Top journals might have a kind of new responsibility to collect and to save each discussion chain and showcase in front of readers.

In addition, the discussion chain system determines the "tendency" of the scientific truth of each fact and is inherently more acceptable than the current binary judgment system, such as a decision of retract. The binary system, more or less, makes authors defensive about their research and provides some incentive to hide the weaknesses of their research from reviewers or readers. Abandoning the binary system and adopting a discussion chain system would be ideal from the perspective of relieving authors from the stress that comes from keen discussions, especially the discussion on maintaining their careers. The application of generative AI to the evaluation discussion chain, not limited to the citation relationship but placing greater priority on the range of criticism, must reduce the evaluation cost.

Now is the time to consider the value of the chain of discussion that is woven by scholars in the historical and international traditions. We must serve the system by respect and funding. We also welcome new members and keep them in the system. Discussion chain–based methodology can protect scientists from the lure of misconduct and questionable research practices. Small errors or mysterious results must be published, and it regarded as a good practice to encouraged others to elongate discussion chain. Moreover, it enables us to calmly conduct science practices and concentrate on science by reducing the noise of nonessential evaluations, which are generally numerical ones. This direction would be acceptable for research communities after the San Francisco Declaration on Research Assessment DORA[8], avoiding overutilization of metrics to evaluate scientists themselves and their promotion and funding. We primarily just maintain the chain of discussion to uncover the truth.

**Acknowledgments**

The author would like to thank Enago (www.enago.jp) for the English language review and to thank to frank discussion among Kumano Dormitory community and Graduate School of Biostudies and KUINS at Kyoto University in Japan.